\documentstyle[12pt]{article}
\newcommand{\be}{\begin{equation}}
\newcommand{\ee}{\end{equation}}
\newcommand{\bea}{\begin{eqnarray}}
\newcommand{\eea}{\end{eqnarray}}
\newcommand{\ba}{\begin{array}}
\newcommand{\ea}{\end{array}}
\newcommand{\beas}{\begin{eqnarray*}}
\newcommand{\eeas}{\end{eqnarray*}}
\newcommand{\bes}{\begin{equation*}}
\newcommand{\ees}{\end{equation*}}

\newcommand{\nn}{\nonumber}

\newcommand{\f}{\frac}
\newcommand{\dagg}{\dagger}

\def\det           {\mbox{\rm det}\,}

\def\i2           {\mbox{$\frac{i}{2}$}}

\def\al           {\alpha}

\def\alb           {{\bar{\alpha}}}
\def\bet           {\beta}
\def\beb           {{\bar \beta}}

\def\ch           {\chi}
\def\chb          {\bar \chi}

\def\del           {\delta}

\def\ep           {\epsilon}
\def\epb           {{\bar \epsilon}}
\def\vep           {\varepsilon}
\def\et           {\eta}

\def\etb           {{\bar \eta}}

\def\ga           {\gamma}
\def\gab           {{\bar \gamma}}

\def\Ga           {\Gamma}

\def\ph           {\phi}

\def\ps           {\psi}
\def\psb          {\bar {\psi}}
\def\chb          {\bar {\chi}}
\def\rh           {\rho}

\def\si           {\sigma}
\def\sib          {{\bar \sigma}}

\def\th{\theta}

\def\alb           {{\bar{\alpha}}}

\def \Qt {{\tilde Q}}

\def\rhb{{\bar {\rh}}}

\begin{document}

\title{ On supermembrane actions on Calabi-Yau 3-folds }
\author{ A. Imaanpur \thanks{Email: aimaanpu@theory.ipm.ac.ir} \\
{\small {\em Department of Physics, School of Sciences,}}\\
{\small {\em Tarbiat Modares University, P.O.Box 14155-4838, 
Tehran, Iran, and}}\\
{\small {\em Institute for Studies in Theoretical Physics and Mathematics}} \\
{\small {\em P.O.Box 19395-5531, Tehran, Iran}}}

\maketitle
\begin{abstract}
 In this note we examine the supermembrane action on Calabi-Yau 
3-folds. We write down the Dirac-Born-Infeld part of the action, and 
show that it is invariant under the rigid spacetime supersymmetry.

\end{abstract}

\section{Introduction}

The study of $p$-branes as solitonic solutions of supergravity has now 
become indispensable for the understanding of nonperturbative effects 
in superstring theory. 
On flat spacetimes $p$-branes are described by an action 
which consists of two parts; the Dirac-Born-Infeld (DBI) part, and the 
Wess-Zumino part. The latter can only be written down for some specific 
spacetime dimensions. For supermembranes, for instance, it exists only for 
4, 5, 7, and 11 spacetime dimensions \cite{TOW}.  
Both parts of the action are invariant under 
the rigid spacetime supersymmetry transformations, however, for 
a specific choice of normalization 
of the Wess-Zumino term, the whole action turns out to have an extra 
local symmetry known as $\kappa$-symmetry. 
The $\kappa$-symmetry allows one to remove 
the redundant fermionic degrees of freedom leaving only the physical 
ones. 

$p$-branes (and also D-branes) have so far been 
studied  on flat spacetimes, and in some cases on $AdS$ spaces 
(look at \cite{TOW, PD} and the references therein, for instance).  
On Calabi-Yau manifolds, on the other hand, the action is not known, 
though, in some cases they have been studied through their low 
energy effective action \cite{BSV, BBS}, which for D$p$-branes is just 
the dimensional reduction of 10d super Yang-Mills theory down to $p +1$ 
dimensions \cite{WB}. Along these lines, an effective action of D-branes 
on Calabi-Yau 3-folds was also introduced in \cite{ALI}.

The study of $p$-branes on Calabi-Yau 3-folds becomes crucial if we 
are to learn about the nonperturbative effects of superstring 
theory or M-theory compactified on 
such manifolds. So it is of importance to find the effective action 
of $p$-branes on Calabi-Yau manifolds. This note is an attempt to construct 
the action of supermembranes on Calabi-Yau manifolds. In section 2, we begin 
with  the preliminaries. The field decompositions on Calabi-Yau 
3-folds are used to write down the supersymmetric invariant 
one-forms $\Pi$. However, these forms will not 
be invariant under supersymmetry transformations as soon as we 
lift them on Calabi-Yau  manifolds. We then show  that 
the supersymmetry transformations 
on a Calabi-Yau 3-fold can be deformed to look like the 
BRST transformations. This fact is then used 
to construct the supersymmetric DBI action of supermembranes.

\section{Preliminaries and the construction of the action}
Consider a Calabi-Yau 3-fold $M$ which has the holonomy group $SU(3)$. 
Take $\th$ to be the singlet spinor on $M$ with the negative chirality. 
Then, as in \cite{JAJ, ALI}, 
we can use the following Fierz identities
\bea
\th\th^{\dagg} + \f{1}{2}\ga^\alb\th^*\th^t\ga_\alb =\f{1}{2}(1-\ga_7)\\
\th^*\th^{t} + \f{1}{2}\ga^\al\th\th^\dagg\ga_\al =\f{1}{2}(1+\ga_7)
\eea
to decompose a complex spinor $\Psi =\Psi_L +\Psi_R$ on $M$. 
Here we have used $\al ,\bet ,\ga ,\ldots $ to represent the complex 
tangent indices on $M$. This results in 
\bea
\Psi_L =\th\psi +\f{1}{2}\ga^\alb\th^*\psi_\alb \Rightarrow 
{{\Psi}^\dagger}_L=\psb\th^\dagg +\f{1}{2}\psb_\al\th^t\ga^\al \nn \\
\Psi_R =\th^*\ch+\f{1}{2}\ga^\al\th\ch_\al \Rightarrow 
{{\Psi}^\dagger}_R =\chb\th^t +\f{1}{2}\chb_\alb \th^\dagg\ga^\alb 
\label{dec} \, ,
\eea
where we have defined 
\bea
\ps = \th^\dagger\Psi_L\ \ \ ,\ \ \ \ps_\alb =\th^t\ga_\alb\Psi_L \nn \\
\ch = \th^t\Psi_R \ \ ,\ \ \ \ch_\al =\th^\dagger\ga_\al\Psi_R \, .\nn
\eea
As $\th$ is a singlet, we have chosen $\ga_\al\th =0$. The covariantly 
constant forms that can be constructed from $\th$ are the K\"ahler 
2-form $k_{\al\beb}=i\th^\dagg\ga_{\al\beb}\th$, and the holomorphic 
3-form $C_{\al\bet\ga}= \th^\dagg\ga_{\al\bet\ga}\th^*$. 

As mentioned in the introduction, the supermembrane action 
can exist in seven dimensions.  
Therefore we first write the action on flat ${\bf R}^7$ spacetime  
\cite{TOW} using the above decomposition of fields, then we lift it 
on to $M\times {\bf R}^1$, where ${\bf R}^1$ represent the time direction. 

Let us first recall the construction of the action 
on flat spacetime \cite{TOW}. Let $m, n, \ldots = 0, 1, 2, \ldots , 6 $ 
represent the tangent indices on the whole manifold, and $\mu ,\nu , \ldots 
=1, 2, \ldots , 6 $ indicate the indices on $M$. $i,j,k,\ldots =1,2,3 $ 
will be the tangent indices on the worldvolume of the brane. 
As for the gamma matrices 
we take
\[
\ga_6 =\ga_0 \ga_1 \ldots \ga_5 \ , \ \ \ \ \ga_7 =i\ga_0 \, ,
\]
where $\ga_\mu$'s are hermitian and $\ga_0^\dagg= -\ga_0$. 
Define 
\[
\Pi^m = dX^m - i{\bar \Psi}\ga^m d\Psi +i d{\bar \Psi}\ga^m \Psi \, ,
\]
where ${\bar \Psi} =\Psi^\dagger \ga_0$. $\Pi^m $ is invariant under the following 
rigid supersymmetry transformations
\be
\del X^m =i{\bar {\vep}} \ga^m \Psi -i {\bar \Psi}\ga^m \vep 
\ \ ,\ \ \ \del\Psi =\vep\, . 
\label{trans1}
\ee
Here $\vep =\vep_L + \vep_R$ is a constant complex six-dimensional spinor. 
On $M$, $\vep$ decomposes just like $\Psi$, however, as 
we are interested in the 
scalar part of the transformations, we drop the triplet part of the $\vep$;
\be
\vep_L =\th\ep \ \ ,\ \ \ \ \vep_R=\th^*\et \, .\label{dec2}
\ee
 
Using the field decompositions in (\ref{dec}) and (\ref{dec2}), 
the supersymmetry 
transformations (\ref{trans1}) read (with $\ph \equiv X^0$)
\bea
&& \del X^\al = \etb\ps^\al + \ep\chb^\al \nn \\
&& \del X^\alb = -\et \psb^\alb - \epb\ch^\alb \nn \\
&& \del \ph = i\epb\ps +i \ep\psb +i \etb\ch +i \et\chb \nn \\
&& \del\psi =\ep \ ,\ \ \ \ \del\ch=\et \nn \\
&& \del\ps^\al =0 \ ,\ \ \ \ \del\ch^\alb=0 \, . \label{trans}
\eea
For the sake of simplicity, in the following, we set $\et =0$ 
in the above transformations and work with just one of the 
supersymmetries survived on $M$. 

The components of $\Pi$, on the other hand, are
\bea
\Pi^\al = dX^\al - \chb D\ps^\al - \ps^\al d\chb - \ps D\chb^\al 
- \chb^\al d\ps +\f{1}{4}C^{\al\bet\ga}
(\psb_\bet D\ch_\ga - \ch_\bet D\psb_\ga )  \nn \\
\Pi^\alb =  dX^\alb + \ch D\psb^\alb +  \psb^\alb d\ch +\psb D\ch^\alb  
+ \ch^\alb d\psb  -\f{1}{4}C^{\alb\beb\gab}
(\ps_\beb D\chb_\gab -\chb_\beb D\ps_\gab )   \nn \\
\!\!\!\!\! \Pi^0 = d\ph -i\psb d\ps -i\chb d\ch -\f{i}{2}
g_{\al\beb}\psb^\beb D\ps^\al 
-\f{i}{2}g_{\al\beb}\chb^\al D\ch^\beb \nn \\ 
-i\ps d\psb -i\ch d\chb -\f{i}{2}g_{\al\beb}\ps^\bet D\psb^\alb 
-\f{i}{2}g_{\al\beb}\ch^\alb D\chb^\bet   \nn \, ,
\eea
where $D\ps^\al = d\ps^\al + dX^\bet \Ga^\al_{\bet\ga}\ps^\ga $.
Further define
\be
M_{ij} = \del_{mn} \Pi^m_i \Pi^n_j \, .\label{M}
\ee
Now as $\Pi$'s are invariant under the SUSY transformations on flat 
spacetime, the action
\be
S_{DBI} = -T\int d^3\si \sqrt {\det M_{ij}}\, ,\label{action}
\ee
is also trivially invariant. 

On a Calabi-Yau 3-fold, however, 
neither $\Pi$ nor the metric is invariant under the SUSY transformations. 
Therefore $M_{ij}= g_{mn}\Pi^m_i\Pi^n_j$ will not be invariant;
\be
\del (g_{mn}\Pi^m_i \Pi^n_j ) =\Delta (g_{mn}\Pi^m_i \Pi^n_j )
= g_{mn}(\Delta \Pi^m_i)\Pi^n_j + g_{mn}\Pi^m_i \Delta \Pi^n_j \, ,
\label{last} 
\ee
where the covariant variation is defined by
\[
\Delta \Pi^\al = \del \Pi^\al +\del X^\rho \Ga^\al_{\rho\si}\ps^\si\, ,
\]
and the last equality in (\ref{last}) follows as the metric $g_{mn}$ is 
covariantly constant. 

To get a supersymmetric action on $M\times {\bf R}^1$, 
firstly we note that on {\em flat}  
spacetimes the transformations (\ref{trans}) square to zero (BRST-like) 
when acting on any field except $\phi$. Secondly if we define 
the operator $\Qt$  by
\[
\Qt =Q + {\bar Q}\  \ \ , \ \ \ \del = i\ep Q +i\epb {\bar Q}\, ,
\]
we can see that $M_{ij}$ in the action can be written as a BRST-exact term; 
\be
M_{ij} = \del_{mn}\Pi^m_i \Pi^n_j = 
\{ \Qt\ ,\ \  \f{i}{2}(\ps +\psb ) \del_{mn}\Pi^m_i \Pi^n_j \}\, .
\label{MIJ}
\ee
This follows as $\{\Qt\ ,\ \f{i}{2}(\ps +\psb )\} =1$. 
If we could maintain the BRST property of $\Qt$ on $M$, with a 
choice of $M_{ij}$ as in (\ref{MIJ}), it would be straightforward to 
construct the supersymmetric action on a Calabi-Yau 3-fold. 
All we need to do is to compute the term 
\be   
\{\Qt\ ,\ \f{i}{2}(\ps +\psb ) g_{mn}\Pi^m_i \Pi^n_j\} \label{com}
\ee
to get $M_{ij}$ on $M\times {\bf R}^1$. 
The invariance of the action under $\Qt$ then follows as $\Qt^2=0$;
\bea
\{\Qt\, ,\, S_{DBI} \}&=& -\f{T}{2}\int d^3\si \sqrt {\det M_{ij}}\,  M^{ij} 
\{\Qt\, ,\, M_{ij} \} \nn \\
&=& -\f{T}{2}\int d^3\si \sqrt {\det M_{ij}}\, M^{ij}
\{\Qt\, , \{ \Qt\, ,\ \  \f{i}{2}(\ps +\psb ) 
g_{mn}\Pi^m_i \Pi^n_j \} \} =0 \, .
\eea
In the following, we will see how this method works.  
  
First of all, on $M\times {\bf R}^1$ 
we need to covariantize the transformations (\ref{trans}). 
The only transformation 
which needs to change is that of $\ps^\al$. So we write
\[
\del\ps^\al = -\del X^\rho \Ga^\al_{\rho\si}\ps^\si  \, . 
\]
With this change, however, $\del$ does not square to zero anymore when 
acting on $\ps^\al$. To maintain this property of $\del$, we add 
another term proportional to the Riemann tensor as follows
\[
\del\ps^\al = -\ep\chb^\rho \Ga^\al_{\rho\si}\ps^\si 
-i\ep \psb \chb^\bet\ch^{\rhb}\ps^\ga R^\al_{\ \bet\rhb\ga} \, , 
\]
or
\[
\Delta \ps^\al =-i\ep\psb\chb^\bet\ch^\rhb\ps^\ga R^\al_{\ \bet\rhb\ga}\, .
\]
Noting that on K\"ahler manifolds $R^\al_{\ \bet\rho\ga}=0$ and 
$R^\al_{\ \bet\rhb\ga}= R^\al_{\ \ga\rhb\bet}$, it is easy to check that 
with this change $\del_{\ep_1}\del_{\ep_2} \ps^\al =0$. 

Taking $M_{ij}$ as in (\ref{com}) we need to work out the variation 
of $\Pi$'s. Since the holomorphic 3-form $C^{\al\bet\ga}$ 
is covariantly constant, we obtain
\bea
&& \Delta \Pi^\al_i = -\chb\Delta (D_i\ps^\al) +\partial_i\chb\Delta\ps^\al 
-\ps\Delta(D_i\chb^\al) 
+\f{1}{4}C^\al_{\ \beb\gab}\left( \Delta\psb^\beb D_i\ch^\gab +\psb^\beb 
\Delta(D_i\ch^\gab) -\ch^\beb\Delta(D_i\psb^\gab)\right) \nn \\
&&\ \ \ \ \ \ \Delta \Pi^0_i =-\f{i}{2}g_{\al\beb}(\Delta\psb^\beb D_i\ps^\al 
+\psb^\beb\Delta (D_i\ps^\al) +\chb^\al\Delta (D_i\ch^\beb)) 
+ {\rm h.c.} \nn \, ,
\eea 
where
\bea
\Delta (D_i\ps^\al) &=& - (\ep\chb^\bet\ps^\si X_i^\rhb 
+\epb\ch^\rhb\ps^\si X_i^\bet )R^\al_{\ \bet\rhb\si}
-i\ep D_i(\psb\chb^\bet \ch^\rhb\ps^\ga R^\al_{\ \bet\rhb\ga}) \nn \\
\Delta (D_i\chb^\al ) &=& -\epb X^\bet _i \ch^\rhb\chb^\ga 
R^\al_{\ \bet\rhb\ga}
\, .
\eea
So finally we find the following supersymmetric action for 
membranes on $M\times {\bf R}^1$
\be
S_{DBI} = -T\int d^3\si \sqrt {\det M_{ij}}\, ,
\ee
with
\bea
M_{ij}& =& g_{mn}\Pi^m_i \Pi^n_j  \nn \\
&-& \f{i}{2}(\ps +\psb) g_{\al\beb} \Pi^\beb_{(j} \left 
[ (i\chb\chb^\bet\ps^\si X_{i)}^\rhb 
+i\chb\ch^\rhb\ps^\si X_{i)}^\bet \right.\nn \\
&+& \partial_{i)}\chb\psb\chb^\bet\ch^\rhb\ps^\si 
+i \ps\ch^\rhb\chb^\si X_{i)}^\bet ) R^\al_{\ \bet\rhb\si}
- \chb D_{i)}(\psb\chb^\bet\ch^\rhb\ps^\ga R^\al_{\ \bet\rhb\ga}) \nn \\
&+& \f{1}{4}C^\al_{\ \beb\gab}\left( (\ps\ch^\sib\chb^\rho\psb^\etb 
D_{i)}\ch^\gab +i \psb^\gab \chb^\rho\ch^\etb X_{i)}^\sib\right.  \nn \\
&-& \left. \left. i\ch^\gab\ch^\sib\psb^\etb X_{i)}^\rho 
-i\ch^\gab\chb^\rho\psb^\etb 
X^\sib_{i)} ) R^\beb_{\ \sib\rho\etb} 
+ \ch^\beb D_{i)}(\ps\ch^\sib\chb^\rho\psb^\etb R^\gab_{\ \sib\rho\etb}) 
\right)\right] \nn \\
&+&\f{i}{4}(\ps +\psb)  g_{\al\beb} \Pi^0_{(j}
\left[ -i\ps\ch^\sib\chb^\rho\psb^\gab 
D_{i)}\ps^\al R^\beb_{\ \sib\rho\gab} \right. \nn \\
&-& \left.\psb^\beb (\chb^\si\ps^\ga X_{i)}^\rhb + 
\ch^\rhb\ps^\ga X_{i)}^\si ) 
R^\al_{\ \si\rhb\ga} -i \psb^\beb D_{i)}(\psb\chb^\bet\ch^\rhb\ps^\ga 
R^\al_{\ \bet\rhb\ga}) \right] + {\rm h.c.} \nn \, .
\eea  
As mentioned earlier, since $\del_{\ep_2}\del_{\ep_1}$ acting on any 
field\footnote{$\del^2$ acting on $\phi$ does not give zero, but this 
does not cause any harm as this field appears in 
$\Pi^0$ as $d\phi$ and $\del^2$ acting on $\Pi^0$ still gives zero.} 
gives zero
it follows that $\Qt^2 =0$, which ensures the invariance of the action under 
$\Qt$.

Apart from the DBI action which was constructed above, supermembranes 
action on flat spacetime has another part, the Wess-Zumino term. 
One way to obtain the WZ term is to vary the DBI action with respect to 
the $\kappa$-symmetry transformations. One then looks for a supersymmetric 
(up to a total derivative) 3-form on the worldvolume such that its 
variation under the $\kappa$-transformations cancels the variation of 
DBI action. The whole action $S= S_{DBI} + S_{WZ}$ is then invariant under 
both the supersymmetry and $\kappa$-symmetry transformations. We hope to 
return to these issues in future works.

 \pagebreak

\end{document}